\documentstyle[prl,aps,epsfig]{revtex}
\begin{document}
\draft
%
%
\title{Maxwell-Schr\"{o}dinger Equation for 
Polarized Light and Evolution of the Stokes Parameters}
\author{Hiroshi Kuratsuji and Shouhei Kakigi}
\address{Department of Physics, Ritsumeikan University-BKC, 
         Kusatsu City 525-8577, Japan}
\date{\today}
\maketitle
\begin{abstract} 
By starting with the Maxwell theory of electromagnetism, we study 
the change of polarization state of light transmitting through 
optically anisotropic media. The basic idea is to reduce the Maxwell 
equation to the Schr\"{o}dinger like equation for two levels (or states) 
representing polarization. By using the quantum mechanical technique, 
the density matrix, and path integral, the evolution of the Stokes 
parameters results in the equation of motion for a pseudospin 
representing a point on the Poincar\'{e} sphere. Two typical examples 
relevant to actual experiments are considered; the one gives 
the generalized Faraday effect, and the other realizes an optical 
analog of magnetic resonance. 
\end{abstract}
%
%
The study of propagation of light (or an electromagnetic wave) in 
optical media has long been one of the major subjects in physics. 
We mention, for example, the classic monographs \cite{Landau,Born}, 
and the works on nonlinear optics \cite{Chiao,Swartz}. 
The characteristic quantity describing the light propagation is the 
concept of polarization. The study of polarization has also a long 
history \cite{Landau,Born}, which forms a basis of modern crystal 
optics. The simple way to describe the polarization state is given 
by the Stokes parameters (or vector). Geometrically, the Stokes 
vector is realized as a point on the so-called Poincar\'{e} sphere. 
The Stokes vector or Poincar\'{e} sphere play a powerful role for 
analyzing the change of the polarization state of light transmitting 
through anisotropic optical media \cite{Born}. As for the equation 
for evolution of the Stokes parameters, the phenomenological 
description has been known in the area of optics, which uses special 
mathematical device such as the Jones vector or M\"{u}ller matrix 
 \cite{Azzam,Brosseau,Jones}.

Having given a brief overview of the developments achieved so far, 
we address a novel formalism of the evolution for the polarization 
state of light transmitting through anisotropic media. Apart from the 
previous phenomenological approaches \cite{Azzam,Brosseau,Jones}, our 
theory is based on the first principle starting from the Maxwell 
theory of electromagnetism \cite{Landau}, where we use the more 
refined form than the original Maxwell equation. Namely, the Maxwell 
equation is reduced to the wave equation \'{a} la Schr\"{o}dinger 
equation for two levels \cite{foot}, which is of first order 
in time (we call this the Maxwell-Schr\"{o}dinger equation hereafter) 
and the dielectric tensor plays a role of Hamiltonian. 
By applying the technique used in usual quantum mechanics, such as 
density matrix as well as path integral, to the Maxwell-Schr\"{o}dinger 
equation, we obtain the evolution equation for the Stokes parameters 
as an equation for a pseudospin which represents a point on 
the Poincar\'{e} sphere. 
This is our main consequence. As typical applications of the equation 
of motion for pseudospin, we consider the polarization change in 
specific media, for which the dielectric tensor has the same structure 
as the Hamiltonian for a real spin in external magnetic field. 
Specifically we are concerned with two cases: The first example is 
the pseudospin in uniform ``magnetic field'' which leads to the 
generalized Faraday effect.
The second example is the pseudospin in oscillating as well as 
uniform field, by which we conjecture a possible occurrence of 
an optical analog of the nuclear magnetic resonance (NMR). 
These examples may be accessible to actual experiments and would 
enable us to reveal new aspects of polarization phenomena that 
have not been expected by previous works.

\paragraph*{Maxwell-Schr\"{o}dinger equation.---} 
We consider the plane electromagnetic (EM) wave of the wave vector 
$k$ ($k$ means the wave vector in the vacuum) travelling through 
the dielectric medium in the $z$ direction. 
The medium is anisotropic with respect to the propagation direction 
and let $\hat \epsilon$ be the dielectric tensor. 
We assume $z$ axis to be one of the principal axes of the dielectric 
tensor, namely, the direction along which the one of the eigenvalue of 
$\hat \epsilon$. When the medium is isotropic, the eigenvalue is 
prescribed to take the value $n_0$. 
Thus the EM wave has the form like 
$D(z,t) = {\bf D}(z)\exp[i\omega t], $ and the wave equation for 
the displacement vector 
$ {\bf D}(z) $ is given by \cite{Landau}
\begin{equation}
{d^2 {\bf D} \over dz^2} + k^2 \hat \epsilon{\bf D} = 0 , 
\end{equation}
where $k = {\omega \over c}$. In the geometry under consideration, 
the dielectric tensor is taken to be $2 \times 2$ matrix. Under the 
most general condition that is governed by the external static 
electric and magnetic fields or mechanical constraint, $\hat\epsilon$ 
can be chosen to be a Hermitian matrix \cite{Landau}, which means that 
the medium is transparent for the light transmission (we will consider 
elsewhere the case that there is an effect of absorption of light). 
Furthermore we consider the general situation that the medium is 
inhomogeneous, namely, the $\hat \epsilon$ depends on $z$. 
We now set the ansatz for the wave ${\bf D}$ 
\begin{equation}
{\bf D}(z) = {\bf f}(z)\exp[ikn_0z] , 
\end{equation}
where the amplitude ${\bf f}(z)$ is given by the $2 \times 1$ row vector, 
\begin{equation}
{\bf f} = {}^t(f_1, f_2) = f_1 {\bf e_1} + f_2{\bf e_2} . 
\end{equation}
{\bf f} is a slowly varying function of $z$ compared with the wave 
length, namely, $k >> \vert f' \vert$, which implies that we consider the 
short range approximation. Here $({\bf e}_1, {\bf e}_2)$ denotes the 
basis of linear polarization. In the short wave approximation, the 
amplitude is shown to satisfy the equation
\begin{equation}
i\lambda{d {\bf f} \over dz} + (\hat {\epsilon}-n_0^2){\bf f} = 0 , 
\end{equation}
where $\lambda \equiv {n_0 \over k} $ is just the wave length in 
the medium of refractive index $n_0$ divided by $2\pi$. Note that the 
second order differential term $f''$ is discarded, since this is much 
smaller than the first order differential $f'$. In this way, the 
above equation can be regarded as an analog of the Schr\"{o}dinger 
equation for two-level state, where $\lambda$ just plays a role of 
the Planck constant and $z$ plays a role of the time variable. 
The components $(f_1, f_2)$ of the vector ${\bf f}$ couple each other 
to give rise to the change of polarization and the ``Hamiltonian'' is 
given by $\hat h = \hat\epsilon-n_0^2$. This form of $\hat h$ 
represents the deviation from the isotropic value, 
that is, ``degree of anisotropy'', namely, the deviation governs 
the change of polarization state. From the hermiticity, the most 
general form of this is written as 
\begin{equation}
\hat h = \left(
  \begin{array}{cc}
   \alpha          & \beta + i\gamma \\
   \beta -i\gamma  & -\alpha  
  \end{array}
\right) . 
\end{equation}
Now for the later use, it is convenient to transform the basis of 
linear polarization ${\bf f}$ into the circular basis \cite{Sakurai}, 
that is, ${\bf e_{\pm}}={1 \over \sqrt 2}({\bf e}_1 \pm i{\bf e}_2)$, 
hence the Schr\"{o}dinger equation becomes 
\begin{equation}
i\lambda{d \psi \over dz} = \hat H \psi , 
\end{equation}
where $\psi = T{\bf f}= {}^t(\psi_1^{*}, \psi_2^{*}), \; \hat H = 
T \hat h T^{-1}$. Here $T$ is the unitary transformation of 
the $2 \times 2$ matrix given by 
\begin{equation}
 T = {1 \over \sqrt 2} 
\left(\begin{array}{cc}
        1 &  i \\
        1 & -i 
      \end{array}
\right) . 
\end{equation}
Thus the transformed Hamiltonian turns out to be 
\begin{eqnarray}
\hat H = T \hat h T^{-1} = 
\left(\begin{array}{cc}
                \gamma         &  \alpha + i\beta \\
                \alpha -i\beta & -\gamma 
      \end{array}
\right) , 
\label{transformedh}
\end{eqnarray} 
which is written in terms of the Pauli spin; $\hat H = 
\sum_{i=1}^3h_i\sigma_i$. The formal solution of the above 
Schr\"{o}dinger equation is given by $\psi(z) = \hat T(z) \psi(0)$
with $\hat T(z)$ being the evolution operator, 
\begin{equation} 
\hat T(z) = P\exp[-{i\over \lambda}\int \hat H(z)dz] . 
\label{evolution}
\end{equation}
Here $P$ denotes the path ordered product which is necessary to handle 
the $z$ dependence of $\hat H$.

\paragraph*{Density matrix and equation of motion of the Stokes 
parameters.---} 
We now consider the reduction of the above Schr\"{o}dinger equation. 
This is carried out by using the density matrix, for which we have 
two cases, the pure polarization and the mixed polarization 
(or partially polarized) state. Here we restrict ourselves to the 
former case in order to simplify the argument \cite{Landau2}, 
thus the density matrix is defined as 
\begin{equation}
\rho =  \psi\psi^{\dagger} = \left(
  \begin{array}{cc}
   \psi_1^{*}\psi_1 & \psi_1\psi_2^{*} \\
   \psi_2\psi_1^{*} & \psi_2^{*}\psi_2 
  \end{array}
 \right) . 
\end{equation}
 In terms of the components of the function $\psi$ given above, we 
have the definition for the Stokes parameter \cite{Wick}; 
$S_i = \psi^{\dagger}\sigma_i\psi, S_0 = \psi^{\dagger}1\psi$, 
where $\sigma_i (i = 1, 2, 3)$ means the Pauli spin matrix. 
These variables satisfy the relation $ S_0^2 = S_1^2 + S_2^2 + S_3^2$, 
which is equivalent to the equation ${\rm det}\rho = 0$. 
Furthermore, in the case that the Hamiltonian is Hermitian, 
we can adopt the conservation of 
probability $\psi_1^{*}\psi_1 + \psi_2^{*}\psi_2 = 1$. So if we 
use the spinor parametrization 
\begin{equation}
\psi_1 = \cos{\theta \over 2} , \quad 
\psi_2 = \sin{\theta \over 2}   \exp[i\phi] , 
\end{equation}
we have $\rho = {1\over 2}(1 + {\bf S}{\bf \sigma})$, 
where the vector ${\bf S} = (S_x, S_y, S_z)$ is given by 
\begin{equation}
S_x = \sin\theta \cos\phi , \quad 
S_y = \sin\theta \sin\phi , \quad 
S_z = \cos\theta . 
\end{equation}
This forms the Stokes vector and is described by the point 
on the Poincar\'{e} sphere. We illustrate some typical values: 
(i) $\theta=0$; the north pole that corresponds to 
the left handed circular polarization. 
(ii) $\theta=\pi$; the south pole that corresponds to 
the right handed circular polarization. 
(iii) $\theta={\pi \over 2}$; the equator which represents 
the linear polarization. The equation of motion for 
the density matrix is written as 
\begin{equation}
i\lambda{d\rho \over dz} = [\hat H, \rho] . 
\label{density}
\end{equation}
Here $[\hat H, \rho] \equiv \hat H\rho-\rho \hat H$. 
Using the commutation relation for the Pauli spin, 
$[\sigma_i, \sigma_j] = 2i\epsilon_{ijk}\sigma_k$, 
we can deduce the equation of motion for the pseudospin 
from (\ref{density})
\begin{equation}
\lambda{d{\bf S} \over dz} = {\bf S}\times {\bf G} , 
\end{equation}
where the effective ``magnetic field'' is defined as 
${\bf G} = (2\alpha, 2\beta, 2\gamma)$. 
If we introduce the ``classical'' counterpart of the Hamiltonian 
(\ref{transformedh})
\begin{eqnarray}
H & = & 2\alpha S_x + 2\beta S_y + 2\gamma S_z \quad \nonumber \\
  & = & 2\alpha \sin\theta\cos\phi + 
  2\beta \sin\theta\sin\phi + 2\gamma\cos\theta ,
\end{eqnarray}
we have an alternative form of the equation of motion \cite{Berezin}
\begin{equation}
\dot\theta = {1\over \lambda\sin\theta}
             {\partial H \over \partial \phi} , \quad 
\dot\phi   =-{1 \over \lambda\sin\theta}
             {\partial H \over \partial \theta} . 
\label{angle}
\end{equation}

The equation of motion (\ref{angle}) can also be obtained as a result 
of the asymptotic limit of $\lambda \rightarrow 0$ \cite{note}. 
This may be achieved by the fact that the set of states for 
pseudospin, ${\psi \equiv \vert \psi>} $ forms a Bloch state that 
satisfies the completeness relation: 
$\int \vert \psi>d\mu<\psi \vert = 1$ with measure 
$d\mu = \sin\theta d\theta d\phi$ (just the volume on the Poincar\'{e} 
sphere). Let us consider the transition amplitude that is given by 
sandwiching the evolution operator (\ref{evolution}) with two initial 
and final spin states. By adopting the procedure of ``time slicing'' 
and inserting the completeness relation at each time division, we get 
the path integral expression \cite{Suzuki}
\begin{equation}
  <\psi_f\vert \hat T(z)\vert \psi_i> 
 = \int \exp \left\{ {i\over \lambda}S[\psi^{*}, \psi]\right\}D\mu(\psi) 
\end{equation}
with the path measure $D\mu(\psi)\equiv \prod_z d\mu[\psi^{*}(z), 
\psi(z)]$ and $S$ is the ``action function''
\begin{equation}
 S= \int <\psi \vert i\lambda {d \over dz}-\hat H \vert \psi> dz . 
\end{equation}
In the limit of $\lambda \rightarrow 0$, we have the stationary phase 
condition $\delta S = 0$ leading to the equation of motion for 
the pseudospin, i.e., (\ref{angle}).

\begin{figure}[htpb]
\begin{minipage}[b]{8.2cm}
\epsfig{file=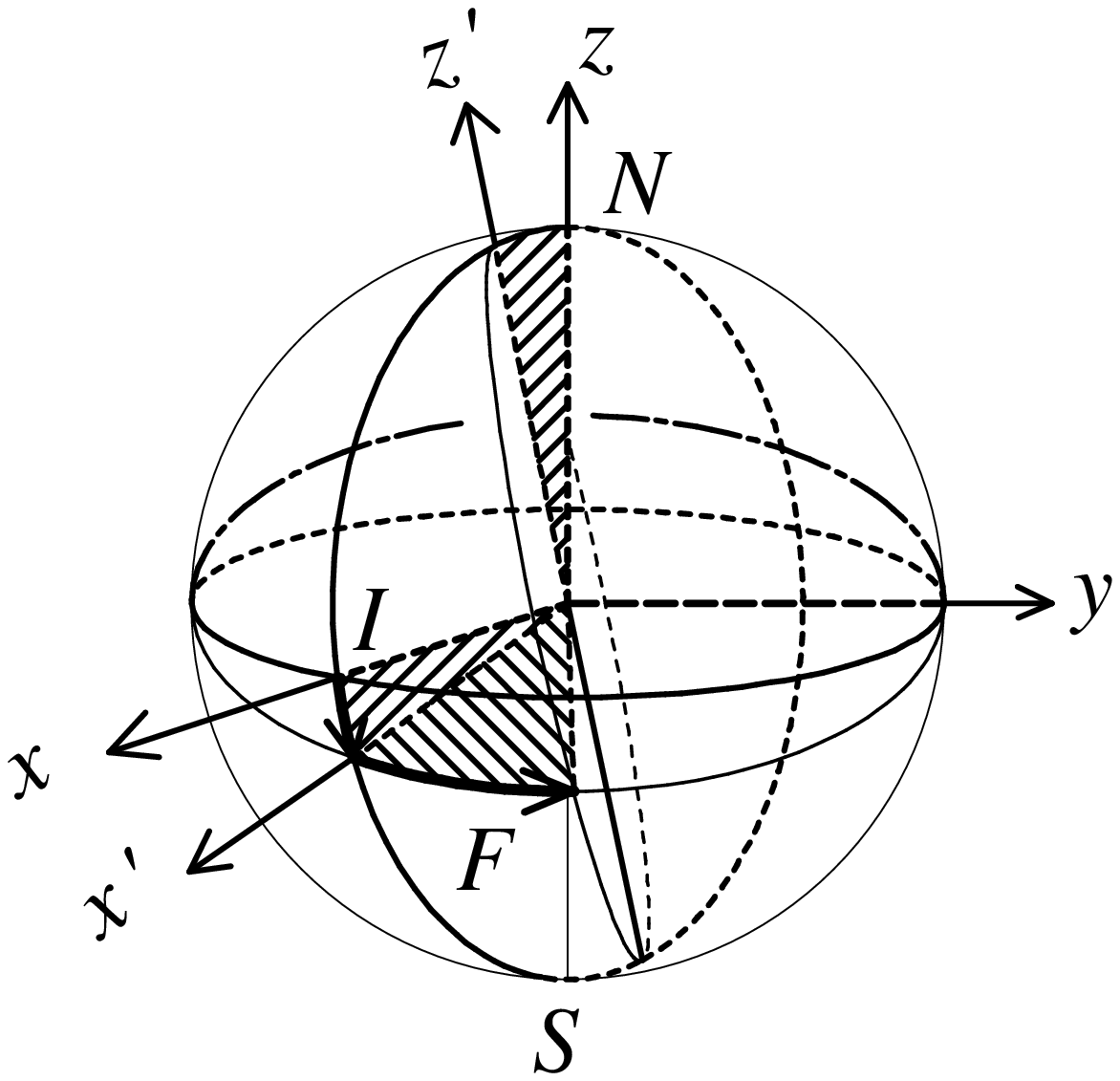,width=8.2cm}
\caption{ The initial linear polarization, which is marked by $I$, 
changes to the final elliptic polarization are marked by $F$.}
\label{fig1}
\end{minipage}\hfill
\begin{minipage}[b]{8.2cm}
\epsfig{file=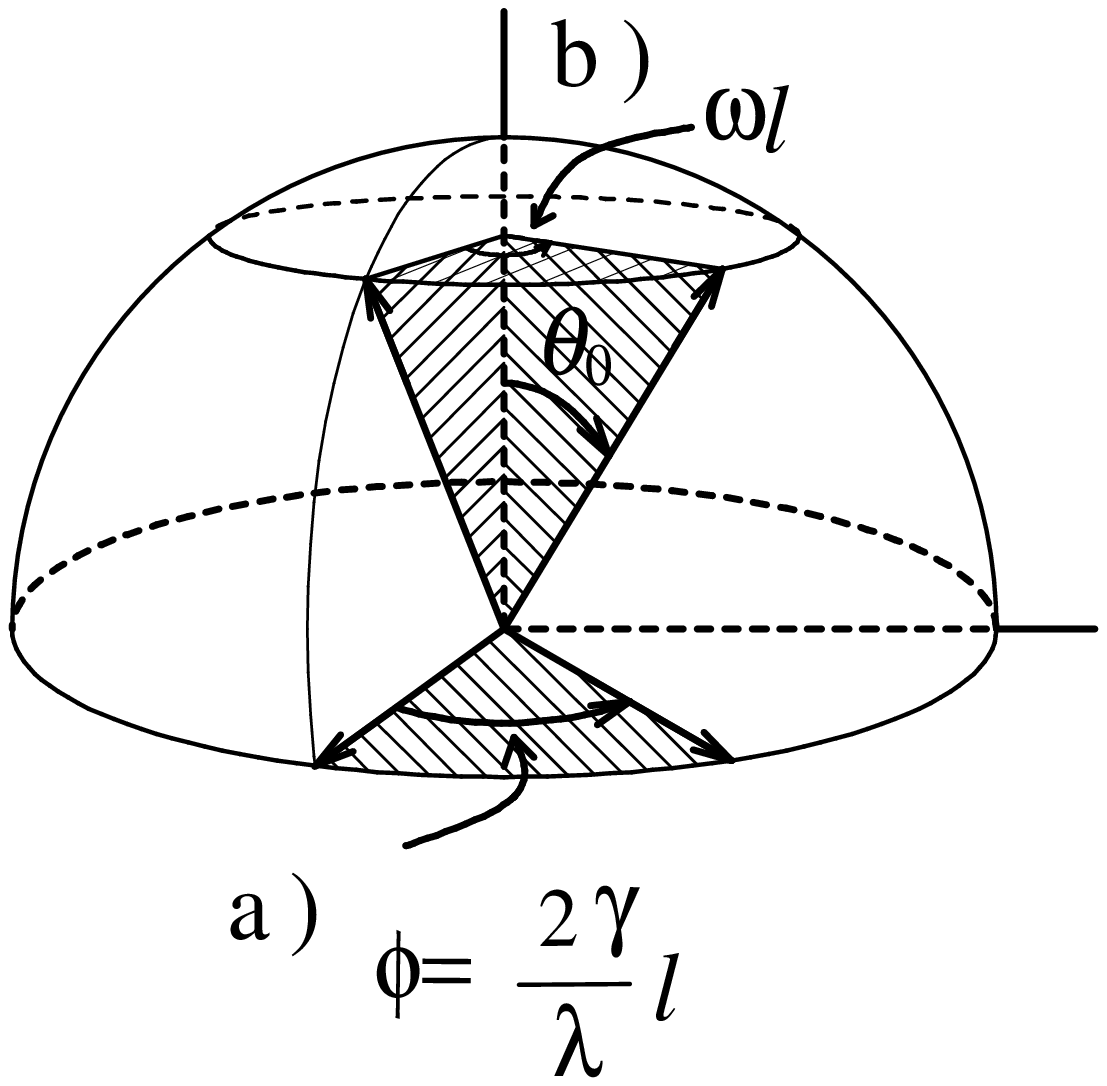,width=8.2cm}
\caption{({\it a}) The change of linear polarization by an amount of 
$\phi = {2\gamma \over \lambda}l$, \, ({\it b}) The change of the axis 
of elliptic polarization by an amount of $\omega l$.}
\label{fig2}
\end{minipage}
\end{figure}

\paragraph*{Typical applications.---} 
We shall consider some special cases that can be described by 
the general formalism. (i) We first consider the model for which 
the dielectric tensor depends on the external magnetic field 
as well as electric field. The kinematical symmetry implies that 
$\hat h$ has the form
\begin{equation}
\hat h = \left(
  \begin{array}{cc}
    \alpha  & i\gamma \\
  -i\gamma  & -\alpha  \\ 
  \end{array}
 \right) . 
\end{equation}
Here $\gamma$ is proportional to the uniform magnetic field (the 
strength is ${\bf B}$) applied in the $z$ direction; $\gamma = {\rm g}B$ 
 \cite{foot2}. Thus according to the formula (\ref{transformedh}), 
it is transformed to $\hat H = \gamma\sigma_z + \alpha \sigma_x$. 
This is further transformed to the Hamiltonian for the spin in 
uniform field of strength $\Gamma$, that is applied in the 
$z'$ direction by rotating about the $y$ axis by an amount of 
the angle $\eta$, such that 
\begin{equation}
S_x' = \cos\eta S_x - \sin\eta S_z , \quad 
S_z' = \sin\eta S_x + \cos\eta S_z , 
\end{equation}
where $\Gamma = \sqrt{\gamma^2 + \alpha^2}$ together with the angle 
$\cos\eta = \gamma/ \Gamma$. 
Thus the equation of motion for the pseudospin becomes 
\begin{eqnarray}
{dS_x' \over dz} =  {2\Gamma \over \lambda}S_y , \quad 
{dS_y  \over dz} = -{2\Gamma \over \lambda}S_x'
\end{eqnarray}
for which we get the solution 
$S_x' = \sin({2\Gamma \over \lambda}z + \theta_0), \, 
S_y   = \cos({2\Gamma \over \lambda}z + \theta_0)$ 
and $S_z' = 0$. In terms of the original spin, it gives 
\begin{equation}
S_x =  \cos\eta \sin({2\Gamma \over \lambda} z + \theta_0) , \quad 
S_z = -\sin\eta \sin({2\Gamma \over \lambda} z + \theta_0) . 
\end{equation}
This means the following feature, if the light wave is initially 
in the linear polarization, then it turns out to be an elliptic 
polarization after transmitting through the medium (see Fig. \ref{fig1}). 
As a special case of $\alpha = 0$, the initial polarization changes 
to $S_x = \sin({2\gamma \over \lambda}z + \theta_0), \; 
S_y =\cos({2\gamma \over \lambda}z +\theta_0)$, namely, the initial 
polarization plane simply changes by an amount of the angle 
$\phi = {2\gamma \over \lambda}l$ after travelling the distance $z = l$. 
This rotation of the polarization plane is known to be 
the Faraday effect [see Fig. \ref{fig2}(a)]. 

(ii) We consider another example for which the polarized light 
propagates in the medium such that the dielectric tensor has a 
periodical structure besides the effect of the external magnetic 
field as in the case (i). For such a system the tensor $\hat h$ is 
given by 
\begin{equation}
\hat h = \left(
  \begin{array}{cc}
  \gamma_0\cos\omega z &  \gamma_0\sin\omega z \\ 
  \gamma_0\sin\omega z & -\gamma_0\cos\omega z 
  \end{array}
  \right)
+ \left(
  \begin{array}{cc}
         0  &  i\gamma \\
   -i\gamma &  0 
  \end{array}
  \right) , 
\label{x}
\end{equation}
which is transformed to 
$\hat H(z)={\bf G}(z){\bullet}{\bf S}$, 
where the pseudomagnetic field has the component: 
\begin{equation}
{\bf G}(z)=(2\gamma_0 \cos \omega z , 
\; 2\gamma_0 \sin \omega z , \; 2\gamma) , 
\end{equation}
namely, a static field along the $z$ axis plus an oscillating field 
rotating perpendicular to it with the frequency $\omega$. 
This feature is familiar in magnetic resonance. 
Using the classical counterpart of the above 
Hamiltonian, which is given as $ H(z) = {\bf G}(z)\cdot{\bf S} = 
2\gamma_0\sin\theta\cos(\phi-\omega z) 
+ 2\gamma \cos\theta $, the equation of motion 
is derived by using the general formula (\ref{angle})
\begin{eqnarray}
\dot\theta &=& -{2\gamma_0 \over \lambda} \sin(\phi-\omega z) , 
\nonumber \\
\dot \phi  &=& -{2\over \lambda}[ \gamma_0 \cot \theta 
              \cos(\phi -\omega z) - \gamma] . 
\end{eqnarray}
One sees that this form of equations of motion allows a special 
solution 
\begin{equation}
\phi = \omega z , \quad \theta = \theta_0 (= {\rm const}) , 
\label{reso}
\end{equation}
where the following relation should hold among the parameters 
$\theta_0, \gamma, \gamma_0$: 
\begin{equation}
\cot \theta_0 = \left({\gamma\over \gamma_0}- {{\lambda \omega} 
                \over 2\gamma_0}\right) . 
\label{y} 
\end{equation}
Equation (\ref{reso}) may be called the ``resonance'' 
solution, since it corresponds to the solution for the forced 
oscillator. The set of parameters $(\gamma, \gamma_0, \omega)$ 
satisfying (\ref{y}) for a fixed value $\theta = \theta_0$ belong to 
a family of resonance solutions. Indeed, this set of parameters forms 
a surface in the parameter space $(\gamma, \gamma_0, \omega)$, which 
we call the ``invariant surface'' and characterizes the resonance 
condition. 
The condition (\ref{y}) is crucial, since all the quantities on the 
right hand side are given in terms of constants that may be allowed 
to be compared with experiment. The physical meaning of the above 
invariant surface is as follows: If given is an initial elliptic 
polarized wave with the angle $\theta_0$ satisfying the resonance 
condition, then there is no change in its shape during its 
transmission. Only its axis rotates by an amount of $\omega l$ 
that is governed by the period inherent in the dielectric tensor 
[see Fig. \ref{fig2}(b)].

Here we give a remark on possible realization of the periodic 
structure of the dielectric tensor in actual systems. 
The one realization is made by using special materials, such as 
cholesteric liquid crystals \cite{deGennes}. The periodic structure 
is naturally realized by the helix inherent in liquid crystal. 
Indeed, the dielectric tensor can be given by the form of the first 
term of (\ref{x}) \cite{deGennes}. Another realization may be given by 
using the mechanical one, namely, let us consider the elastic body 
under the pressure that is periodically modulated, then this causes 
the periodic modulation of the dielectric tensor according to the 
procedure known in the mechanical-optical effect \cite{Landau}. 
The periodical oscillation of the pressure may be generated, 
for example, by the piezoelectric effect. 
Now having settled the periodic structure, we take into account 
the term coming from the uniform magnetic field that is applied 
along the same direction as the axis of helix. We are reminded of 
the analogy with the NMR; if the initial state is in the left handed 
circular polarization, which corresponds to the state of spin-down, 
the probability for transition to the state of spin-up 
(right handed circular polarization) is given by 
\begin{equation}
P(l \rightarrow r) = {(\gamma_0')^2\over(\Omega -\omega)^2
                  + (\gamma_0')^2}\sin^2{1\over 2}\Delta z , 
\end{equation}
where $\gamma_0' \equiv {\gamma_0 \over \lambda}$ and 
$\Omega = {\gamma \over \lambda}$ and 
$\Delta^2 = (\Omega-\omega)^2 + (\gamma_0')^2$. 
If the magnetic field is chosen so as to synchronize the period 
of the helix, we expect an analogous effect with 
the magnetic resonance. 

This work was inspired by the discussion at seminar class that had 
been instructed by one of the authors (H.K.). The authors thank 
the attendees of the seminar. 
\newpage

\begin{references}
\bibitem{Landau}
L. Landau and E. Lifschitz, {\it Electrodynamics in Continuous Media}, 
Course of Theoretical Physics Vol.8 (Pergamon Oxford, 1968).
\bibitem{Born}
M. Born and E. Wolf, {\it Principle of Optics} 
(Pergamon, Oxford, 1975).
\bibitem{Chiao}
R. Y. Chiao, E. Gamire, and C. H. Townes, 
Phys. Rev. Lett. {\bf 13}, 479 (1964).
\bibitem{Swartz}
G. A. Swartzlander, Jr. and C. T. Law, 
Phys. Rev. Lett. {\bf 69}, 2503 (1992), and references therein.
\bibitem{Azzam}
R. Azzam, J. Opt. Soc. Am. {\bf 68}, 1756 (1978).
\bibitem{Brosseau}
C. Brosseau, Opt. Lett. {\bf 20}, 1221 (1995), and references therein.
\bibitem{Jones}
R. C. Jones, J. Opt. Soc. Am. {\bf 38}, 671 (1948).
\bibitem{foot}
The similar reduction has been previously obtained by Chiao 
{\it et al}.: R. Chiao and J. Goldine, Phys. Rev. {\bf 185}, 
430 (1969), which deals with the nonlinear Schr\"{o}dinger 
equation and does not take into account the anisotropic effect.
\bibitem{Sakurai}
J. J. Sakurai, {\it Advanced Quantum Mechanics} 
(Wiley, New York, 1967).
\bibitem{Landau2}
To consider the partially polarized state means that one should 
treat the time-varying function of the polarization. This implies 
that the polarization is distributed over a certain range of frequency, 
which is denoted by some parameter $\xi$. Thus we treat the partially 
polarized state by using the statistical distribution $P_{\xi}$, 
$\bar\rho =  \sum_{\xi}\psi^{\xi}P_{\xi}\psi^{\xi\dagger}$. 
(Note that the wave function is also paramtetrized by the parameter 
$\xi$). See L. Landau and E. Lifschitz, {\it Classical Fields}, 
Course of Theoretical Physics Vol.5 (Pergamon, Oxford, 1968).
\bibitem{Wick}
G. C. Wick, in {\it Prelude in Theoretical Physics}, edited by 
 A. deShalit, H. Feshbach, and L. van Hove (North-Holland 
 Publishing, Amsterdam, 1966).
\bibitem{Berezin}
see e.g., F. A. Berezin, Commum. Math. Phys. {\bf 40}, 153 (1975).
\bibitem{note}
The procedure given below is also applicable for the case of 
nonlinear media, for which the dielectric tensor is given in terms 
of the quadratic form of the pseudospin. 
\bibitem{Suzuki} 
H. Kuratsuji and T. Suzuki, J. Math. Phys. (N.Y.) {\bf 21}, 419 (1980).
\bibitem{foot2}
It is known \cite{Landau} that the dielectric tensor has the form 
${\bf D} = \epsilon{\bf E} + i{\bf E}\times {\bf \gamma}$ with 
${\bf \gamma} = {\rm g}{\bf B}$ in the case that only the magnetic field 
is applied in the direction of light propagation. 
\bibitem{deGennes}
P. de Gennes, {\it Physics of Liquid Crystal} 
(Oxford University Press, Oxford, 1993),  2nd ed. 
\end{references}
\end{document}